\documentclass[debug,overfull]{epl}
\usepackage{amsmath}

\newcommand\beq{\begin{equation}}
\newcommand\eeq{\end{equation}}
\newcommand\bea{\begin{eqnarray}}
\newcommand\eea{\end{eqnarray}}
\newcommand\non{\nonumber}

\newcommand\al{\alpha}
\newcommand\be{\beta}
\newcommand\si{\sigma}
\newcommand\La{\Lambda}
\newcommand\dg{\dagger}
\newcommand\up{\uparrow}
\newcommand\dn{\downarrow}
\newcommand\bib{\bibitem}

\title{A multi-channel fixed point for a Kondo spin coupled to a junction of
Luttinger liquids}
\author{V. Ravi Chandra \inst{1}, Sumathi Rao \inst{2} \and Diptiman Sen
\inst{1}} 
\institute{\inst{1} Centre for High Energy Physics, Indian Institute of 
Science, Bangalore 560012, India \\
\inst{2} Harish-Chandra Research Institute, Chhatnag Road, Jhusi, Allahabad
211019, India}


\begin{document}

\maketitle

\begin{abstract}
We study a system of an impurity spin coupled to a junction of several
Tomonaga-Luttinger liquids using a renormalization group scheme. 
For the decoupled $S$-matrix at the junction, there is a range 
of Kondo couplings which flow to a multi-channel fixed point 
for repulsive inter-electron interactions; this is associated 
with a characteristic temperature dependence of the spin-flip scatterings. 
If the junction is governed by the Griffiths $S$-matrix, the Kondo couplings 
flow to a strong coupling fixed point where all the wires are decoupled.
\end{abstract}

Although the Kondo effect has been studied for many years and is 
one of the best understood paradigms of strongly correlated systems
\cite{kondo}, it continues to yield new physics, such as in its recent 
manifestations in quantum dots \cite{oreg,rosch}. 
For quantum dots with an odd number of electrons, a Kondo resonance occurs 
at the Fermi level of the leads, which is seen as a peak in the conductance.

If the leads are one-dimensional, inter-electron interactions turn them into
Tomonoga-Luttinger liquids (TLLs); this changes the physics considerably. The
Kondo effect has been studied for a system with two TLL leads 
\cite{lee,furusaki,fabrizio,frojdh} and for crossed TLL wires 
\cite{karyn}. For weak potential scattering, the strong coupling fixed point 
(FP) consists of decoupled TLLs, the conductance vanishes at 
$T=0$ via the usual TLL power law \cite{gogolin}, and a spin singlet is 
formed if the impurity has spin 1/2 (Furusaki-Nagaosa point) \cite{furusaki}.

Motivated by recent experiments which probe the Kondo density of states in 
a three terminal geometry \cite{leturcq}, we will study what happens 
when an impurity spin is coupled to a junction of more than two quantum wires 
which are modeled as TLLs. The junction is characterized by an $S$-matrix
which governs the connections between the wires, and the couplings 
of the Kondo spin are described by a $J$-matrix. (Although the $S$-matrix
formalism for calculating the conductance is strictly valid only for 
non-interacting electrons, we will use it here because the interactions will 
be assumed to be weak and will be treated perturbatively \cite{yue,lal1}).

We obtain the renormalization group (RG) equations for the system, and find 
that the flow of the Kondo couplings depends on the form of the $S$-matrix. At
the point where all the wires are decoupled from each other, we find that for 
a large range of initial values of the Kondo couplings, the system flows to a 
multi-channel FP lying at zero coupling. This FP is associated with spin-flip 
scatterings of the electrons from the impurity spin whose temperature 
dependence will be discussed below. [Note that in the language of the standard
$N$-channel Kondo problem, the scattering matrix $S$ is the $N \times N$ 
identity matrix, and $J_{ij}= J_i \delta_{ij}$; anisotropy between the channels
is introduced by having off-diagonal elements in the $S$ and $J$ matrices]. On
the other hand, at the Griffiths $S$-matrix (defined below), there is no stable
FP for finite values of the Kondo couplings, and the system flows towards
strong coupling in two possible ways. In one case, the impurity is strongly
and antiferromagnetically coupled to the electron spin at the junction.
We then perform an expansion in the inverse
of the Kondo couplings and find that the system is now near the decoupled 
$S$-matrix; hence it flows to the multi-channel FP. In the other case, the 
impurity is coupled strongly and ferromagnetically to the electron spin at 
the origin and antiferromagnetically to the neighboring electron spins;
further analysis then depends on the values of $N$ and the impurity spin 
$\cal S$.

We begin with $N$ semi-infinite wires which meet at a junction, where the 
incoming and outgoing fields are related by an $N\times N$ unitary $S$-matrix.
For an electron which is incoming in wire $i$ ($=1,2,\cdots,N$) with spin 
$\al$ ($= \up,\dn$) and wave number $k$ (defined with respect to the 
Fermi wave number $k_F$), the wave function is given by 
\bea
\psi_{i \al k} (x) &=& e^{-i(k+k_F)x} ~+~ S_{ii} e^{i(k+k_F)x} ~~{\rm on ~
wire~} i ~, \non \\
&=& S_{ji} ~e^{i(k+k_F)x} ~~{\rm on ~wire~} j \ne i ~.
\label{wavefn}
\eea
Here $k$ goes from $- \La$ to $\La$.
The second quantized annihilation operator corresponding to the above wave
function is given by $\Psi_{i \al k} (x) = c_{i \al k} \psi_{i \al k} (x)$.
If an impurity spin is coupled to the electrons at the junction, the 
Hamiltonian is given by
\bea
H_0 ~+~H_{\rm spin} &=& v_F ~\sum_i ~\sum_\al ~\int_{- \La}^\La ~
\frac{dk}{2\pi} ~k~ c_{i \al k}^\dg c_{i \al k} ~ \non\\
& & +~ \sum_{i,j} \sum_{\al,\be} ~\int_{-\La}^\La ~\int_{-\La}^\La ~
\frac{dk_1}{2\pi} ~\frac{dk_2}{2\pi} ~ J_{ij} ~{\vec {\cal S}} \cdot ~
c_{i\al k_1}^\dg ~\frac{{\vec \si}_{\al \be}}{2} ~c_{j\be k_2} ~, 
\label{hspin2}
\eea
where the dispersion has been linearized ($E= v_F k$), $J_{ij}$ is a 
Hermitian matrix, ${\vec \si}$ denotes the Pauli matrices, and we are 
assuming an isotropic spin coupling $J_x = J_y = J_z$ for simplicity.

Next, we consider density-density interactions between the electrons of 
the form 
\beq
H_{\rm int} ~=~ \frac{1}{2} ~\int \int ~dx ~dy ~\rho (x) ~U(x-y) ~\rho (y) ~,
\label{hint1}
\eeq
where the density operator $\rho$ is given in terms of the second quantized 
electron field $\Psi_\al (x)$ $=\sum_i\int dk/(2\pi)\Psi_{i\al k}(x)$ as 
$\rho = \Psi^\dg_\up \Psi_\up +\Psi^\dg_\dn \Psi_\dn$. Writing the electron 
field in terms of outgoing and incoming fields as $\Psi_\al (x) ~=~ 
\Psi_{O\al} (x) ~+~ \Psi_{I\al} (x)$, the Hamiltonian in (\ref{hint1}) takes 
the form 
\bea
H_{\rm int} &=& \int dx \sum_{\al,\be} ~[g_1 \Psi_{O\al}^\dg \Psi_{I\be}^\dg
\Psi_{O\be} \Psi_{I\al}+g_2 \Psi_{O\al}^\dg \Psi_{I\be}^\dg \Psi_{I\be} 
\Psi_{O\al} \non \\
& & ~~~~~~~~~~~~+ ~\frac{1}{2} ~g_4 ( \Psi_{O\al}^\dg \Psi_{O\be}^\dg 
\Psi_{O\be} \Psi_{O\al} + \Psi_{I\al}^\dg \Psi_{I\be}^\dg \Psi_{I\be} 
\Psi_{I\al} )].
\label{hint2}
\eea
The parameters $g_1$, $g_2$ and $g_4$ satisfy some RG equations 
\cite{solyom,yue}, and are given by
\bea
g_2 (L) &=& {\tilde U} (0) ~-~ \frac{1}{2} ~{\tilde U} (2k_F) ~+~ \frac{1}{2}~
\frac{{\tilde U} (2k_F)}{1 ~+~ \frac{{\tilde U} (2k_F)}{\pi v_F} \ln L} ~,
\non \\
g_1 (L) &=& \frac{{\tilde U} (2k_F)}{1 ~+~ \frac{{\tilde U} (2k_F)}{\pi v_F}
\ln L} ~,~~ {\rm and} ~~ g_4 (L) ~=~ {\tilde U} (0),
\label{g124}
\eea
where $L$ denotes the length scale, and $\tilde U$ is the Fourier transform 
of $U$. (We have ignored umklapp scattering). 

The junction $S$-matrix satisfies an RG equation which was derived in Refs. 
\cite{yue,lal1} in the absence of Kondo couplings. We find that the Kondo 
couplings $J_{ij}$ do not affect the RG flows of the $S$-matrix up to second
order in the $J_{ij}$. Since we are mainly interested in the flows of the 
$J_{ij}$, we will assume for simplicity that we are at a FP of the RG
equations for $S_{ij}$. We will study what happens near two particular FPs 
of $S_{ij}$.

We use the technique of `poor man's RG' \cite{anderson,noz1} to derive the 
RG equations for the Kondo couplings $J_{ij}$. (The details will be presented 
elsewhere). To second order in the couplings $J_{ij}$ and the parameters 
$g_a$ (which are given in Eq. (\ref{g124})), we find that 
\bea
\frac{dJ_{ij}}{d \ln L} ~=~ \frac{1}{2\pi v_F} & & \sum_k ~[~ J_{ik} J_{kj} + 
\frac{1}{2} ~g_2 ~ (S_{ij} ~J_{ik} S^*_{ik} + S^*_{ji} J_{kj} S_{jk}) \non \\
& & ~~~~~~~~-~ \frac{1}{2} (g_2 - 2 g_1) ~(J_{ik} S^*_{kk} S_{kj} + S^*_{ki}
S_{kk} J_{kj} )~],
\label{rgj}
\eea

We now consider two possibilities for the $S$-matrix.
The first case is that of $N$ disconnected 
wires for which the $S$-matrix is given by the $N \times N$ identity matrix 
(up to phases). We consider a highly symmetric form of the Kondo coupling 
matrix (consistent with the symmetry of the $S$-matrix) in which all the 
diagonal entries are $J_1$ and all the off-diagonal entries are $J_2$, with 
both $J_1$ and $J_2$ being real. Eq. (\ref{rgj}) then gives
\bea
\frac{dJ_1}{d \ln L} &=& \frac{1}{2\pi v_F} ~[ J_1^2 + (N-1) J_2^2 + 2 g_1 
J_1 ] ~, \non \\
\frac{dJ_2}{d \ln L} &=& \frac{1}{2\pi v_F} ~[ 2 J_1 J_2 + (N-2) J_2^2 -
(g_2 - 2 g_1) J_2 ]~.
\label{disc}
\eea
(For $N=2$ and $g_1 = 0$, Eq. (\ref{disc}) agrees with the results 
in Ref. \cite{fabrizio}). Since $g_1 (L=\infty) = 0$, Eq. (\ref{disc}) has a 
FP at $(J_1, J_2) =(0,0)$. If $\nu \equiv g_2 (L=\infty)/(2 \pi v_F) > 0$ 
(repulsive interactions), a linear stability analysis shows that this FP
is stable to small perturbations in $J_2$. For small perturbations in 
$J_1$, this FP is marginal; a second order analysis shows that it is stable 
if $J_1 < 0$ and unstable if $J_1 > 0$, i.e., it is the usual {\it
ferromagnetic} fixed point which is found for Fermi liquid leads. However,
the approach to the fixed point is quite different when the leads are
TLLs. At large length scales, the FP is 
approached as $J_1 \sim - 1/\ln L$ and $J_2 \sim 1 /L^\nu$. From this, we 
can deduce the behavior at very low temperatures, namely,
\beq
J_1 ~\sim~ -~ 1/(\ln 1/T) ~, ~~{\rm and} ~~~ J_2 ~\sim~ T^\nu ~.
\label{lowt}
\eeq
This is in contrast to the behavior of $J_2$ for Fermi liquid leads, 
i.e., for $g_1 = g_2 = 0$. In that case, Eq. (\ref{disc}) again gives a 
FP at $(J_1,J_2) = (0,0)$, but $J_2$ approaches zero as $1/ (\ln 1/T)^2$.
Note that $J_2$ (which measures the asymmetry between the channels) 
approaches zero faster than $J_1$ for Fermi liquid leads; but for 
TLLs, it goes to zero much faster, i.e., as a power of $T$.


\begin{figure}[htb]
\onefigure[width=6cm]{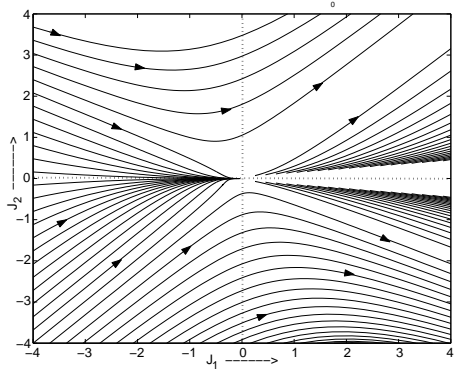}
\caption{RG flows of the Kondo couplings for three disconnected wires, with
${\tilde U} (0) = {\tilde U} (2k_F) = 0. 2 (2 \pi v_F)$.}
\end{figure}

Fig. 1 shows the RG flows for three wires for
${\tilde U} (0) = {\tilde U} (2k_F) = 0. 2 (2 \pi v_F)$.
[This gives a value of $\nu$ which is comparable to what is found in
several experimental systems (see \cite{lal2} and references therein). In 
Figs. 1 and 2, the values of $J_{ij}$ are shown in units
of $2\pi v_F$.] We see that the RG flows take a large range of initial 
conditions to the FP at $(0,0)$. For all other initial 
conditions, we see that there are two directions along which the Kondo 
couplings flow to infinity; these are given by $J_2 = J_1$ and $J_2 = 
- J_1 /(N-1)$ (with $N=3$). This asymptotic behavior can be understood by
analyzing Eq. (\ref{disc}) after ignoring the terms of order $g_1$ and $g_2$.

The second case that we study is called the Griffiths $S$-matrix. Here all the
$N$ wires are connected to each other and there is maximal transmission, 
subject to the constraint that there is complete symmetry between the wires.
The resultant $S$-matrix has, up to phases, all the diagonal entries equal to
$-1 + 2/N$ and all the off-diagonal entries equal to $2/N$. We again consider a
highly symmetric form of the Kondo coupling matrix, with real parameters $J_1$
and $J_2$ as the diagonal and off-diagonal entries respectively. Eq. 
(\ref{rgj}) then gives
\bea
\frac{dJ_1}{d \ln L} &=& \frac{1}{2\pi v_F} ~[J_1^2 + (N-1) J_2^2 ~
+ ~2g_1 ~(1 - \frac{2}{N})^2 ~J_1 ~-~ 4g_1 (1 - \frac{2}{N})~(1 - 
\frac{1}{N}) ~J_2 ~], \non \\
\frac{dJ_2}{d \ln L} &=& \frac{1}{2\pi v_F} ~[ 2 J_1 J_2 +(N-2) J_2^2 ~
- \frac{4g_1}{N} ~(1 - \frac{2}{N}) ~J_1 ~+~ (g_2 - 2 g_1 (1 - 
\frac{2}{N})^2)) ~J_2 ~].
\label{grif}
\eea
(For $N=2$ and $g_1 = 0$, Eq. (\ref{grif}) agrees with the results in Ref.
\cite{furusaki}). Eq. (\ref{grif}) has a FP at the origin (which is unstable 
for $g_2 (\infty) > 0$), and two strong coupling FPs as before. Fig. 2 shows 
a picture of the RG flows for three wires. The couplings are again seen to 
flow to infinity along one of the two directions $J_2 = J_1$ and 
$J_2 = - J_1 /(N-1)$.

\begin{figure}[htb]
\onefigure[width=6cm]{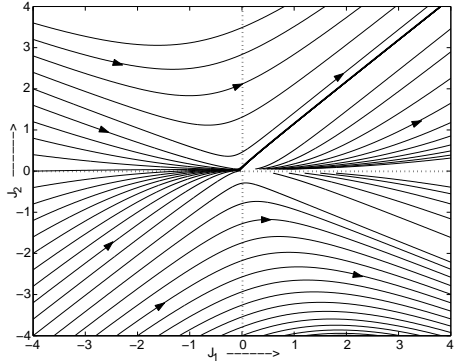}
\caption{RG flows of the Kondo couplings for the Griffiths $S$-matrix for 
three wires, with ${\tilde U} (0) = {\tilde U} (2k_F) = 0. 2 (2 \pi v_F)$.}
\end{figure}

\begin{figure}[htb]
\onefigure[width=5.5cm]{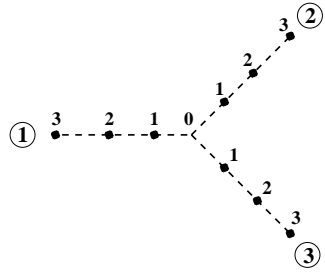}
\caption{A lattice model for the $S$-matrices discussed in the text.}
\end{figure}

We will now see how the different $S$-matrices and RG flows discussed above 
can be interpreted in terms of lattice models as was done for the two-wire 
case in Ref. \cite{furusaki}. The case of $N$ disconnected wires can be 
realized by the lattice model shown in Fig. 3. The Hamiltonian is taken to be
of the tight-binding form, with a hopping amplitude $-t$ on all the bonds, 
except on the $N$ bonds connecting to the junction site $n=0$ where they are 
taken to be zero. (This is equivalent to removing the junction site from the 
system). The impurity spin is coupled to the sites $n=1$ on the different 
wires by the Hamiltonian
\beq
H_{\rm spin} = F_1 ~{\vec S} \cdot ~\sum_i ~\sum_{\al,\be} ~\Psi_\al^\dg 
(i,1) ~\frac{{\vec \si}_{\al \be}}{2} ~\Psi_\be (i,1)~ + ~F_2 ~
{\vec {\cal S}} \cdot ~\sum_{i\ne j} ~\sum_{\al,\be} ~\Psi_\al^\dg (i,1) ~
\frac{{\vec \si}_{\al \be}}{2} ~\Psi_\be (j,1), 
\label{hspin3}
\eeq
where $\Psi_\al (i,1)$ denotes the second quantized electron field at site 1 
on wire $i$. (Eq. (\ref{heff}) below will provide a justification for this 
Hamiltonian). In Eq. (\ref{hspin3}), $F_1$ and $F_2$ denote amplitudes for 
spin-dependent scattering from the impurity within the same wire and between
two different wires respectively.
Now, we find that the Kondo coupling matrix $J_{ij}$ in Eq. (\ref{hspin2}) is 
as follows: all the diagonal entries are given by $J_1$ and all the 
off-diagonal entries are given by $J_2$, where 
\beq
J_1 ~=~ 4 F_1 \sin^2 k_F ~, ~~{\rm and}~~~J_2 ~=~ 4 F_2 \sin^2 k_F
\label{jab1}
\eeq
for modes with redefined wave numbers lying close to zero. The RG flow of
this is given in Eq. (\ref{disc}). In particular, the approach to the 
FP at $(J_1,J_2)=(0,0)$ given by Eq. (\ref{lowt}) at low temperatures implies 
that spin-flip scattering within the same wire or between two different wires 
will have quite different temperature dependences.

The case of the Griffiths $S$-matrix can also be realized by the lattice shown
in Fig. 3 and a tight-binding Hamiltonian. The hopping amplitude is now $-t$ 
on all bonds, except for the $N$ bonds connecting to the junction site where
it is taken to be $t_1 = -t \sqrt{2/N}$. We then find that the $S$-matrix is 
of the Griffiths form for all values of the wave number $k$. The impurity spin
is then coupled to the junction site and the $n=1$ sites by the Hamiltonian
\beq
H_{\rm spin} ~=~ F_3 {\vec {\cal S}} \cdot \sum_{\al,\be} ~\Psi_\al^\dg (0)
\frac{{\vec \si}_{\al \be}}{2} \Psi_\be (0)~ +~ F_4 {\vec {\cal S}} \cdot 
\sum_i \sum_{\al,\be} ~\Psi_\al^\dg (i,1) \frac{{\vec \si}_{\al \be}}{2} 
\Psi_\be (i,1) ~, 
\label{hspin4}
\eeq
where $\Psi_\al (0)$ denotes the electron field at the junction site with spin
$\al$. Then the Kondo coupling matrix $J_{ij}$ in Eq. (\ref{hspin2}) has all
the diagonal entries equal to $J_1$ and all the off-diagonal entries equal to
$J_2$, where
\beq
J_1 ~=~ \frac{4F_3}{N^2} ~+~ 2F_4 ~[~ 1 ~-~ (1 - \frac{2}{N}) \cos 2k_F ~]~,~~
{\rm and} ~~~J_2 ~=~ \frac{4F_3}{N^2} ~+~ \frac{4F_4}{N} \cos 2k_F
\label{j1j2}
\eeq
for modes with wave numbers lying close to zero. Eq. (\ref{grif}) gives the RG
flows of these parameters. Eq. (\ref{j1j2}) implies
\beq
J_1 - J_2 ~=~ 2 F_4 ~(1 - \cos 2k_F)~, ~~{\rm and} ~~~
J_1 + (N-1) J_2 ~=~ \frac{4F_3}{N} ~+~ 2F_4 ~(1 + \cos 2k_F) . 
\label{jab2}
\eeq
Since $0 < k_F < \pi$, $1 \pm \cos 2k_F$ lie between 0 and 2. In the first
quadrant of Fig. 2, we see that $J_1 + (N-1) J_2$ goes to $\infty$ much faster
than $|J_1 - J_2 |$; Eq. (\ref{jab2}) then implies that $F_3$ goes to $\infty$
and $|F_4 | \ll F_3$. In the fourth quadrant of Fig. 2, $J_1 - J_2$ goes to
$\infty$ much faster than $J_1 + (N-1) J_2$; hence $F_4$ goes to $\infty$ and
$F_3$ goes to $-\infty$.
These flows to strong coupling have the following physical interpretations. In
the first case, $F_3$ flows to $\infty$ which means that the impurity spin is
strongly and antiferromagnetically coupled to an electron spin at the junction
site $n=0$; hence those two spins will combine to form an effective spin of 
${\cal S} - 1/2$. In the second case, the impurity spin is coupled strongly 
and ferromagnetically to an electron spin at the site $n=0$, and strongly and 
antiferromagnetically to electron spins at the sites $n=1$ on each of the $N$
wires to form an effective spin of ${\cal S} + 1/2 - N/2$. 

We considered above two kinds of $S$-matrices and found that the Kondo 
couplings flow to infinity for many initial conditions. We will now show 
through an example that the vicinity of a strong coupling FP can be studied 
through an expansion in the inverse of the Kondo coupling \cite{noz1}.
Following the discussion after Eq. (\ref{jab2}), let us assume that the RG 
flows for the case of the Griffiths $S$-matrix have taken us to a strong 
coupling FP along the line $J_2 = J_1$; thus the impurity spin is coupled to
the electron spin at $n=0$ with a large and positive (antiferromagnetic) value
$F_3$, while its couplings to the sites $n=1$ have the value $F_4 =0$. (The 
arguments given below do not change significantly if $F_4 \ne 0$, provided that
$|F_4 | \ll F_3$). From the first term in Eq. (\ref{hspin4}), we see that the
impurity spin couples to an electron at $n=0$ to form an effective spin of 
${\cal S} - 1/2$; the energy of this spin state is $-F_3 ({\cal S}+1)/2$. This
lies far below the high energy states in which an electron at site $n=0$ forms
a total spin of ${\cal S}+1/2$ with the impurity spin (these states have energy
$F_3 {\cal S}/2$), or the states in which the site $n=0$ is empty or doubly 
occupied (these states have zero energy). 

We now perturb in $1/F_3$. The unperturbed Hamiltonian corresponds to $N$ 
disconnected wires along with the impurity spin coupled to the junction site
$n=0$. The perturbation $H_{\rm pert}$ consists of the hopping amplitude $t_1$
on the $N$ bonds connecting the sites $n=1$ to the junction site. Using this
perturbation, we can find an effective Hamiltonian \cite{noz1}. If ${\cal S} 
> 1/2$, we find that the effective Hamiltonian has no terms of order $t_1$, 
and is given by
\bea 
H_{\rm eff} &=& F_{\rm 1, eff} ~{\vec {\cal S}}_{\rm eff} \cdot \sum_i 
\sum_{\al, \be} ~\Psi^\dg_\al (i,1) \frac{{\vec \si}_{\al \be}}{2} 
\Psi_\be (i,1)~ +~ F_{\rm 2, eff} ~{\vec {\cal S}}_{\rm eff} \cdot 
\sum_{i \ne j} \sum_{\al,\be} ~\Psi^\dg_\al (i,1) 
\frac{{\vec \si}_{\al \be}}{2} \Psi_\be (j,1) \non \\
{\rm with} && F_{\rm 1, eff} ~=~ F_{\rm 2, eff} = -~ 
\frac{8 t_1^2}{F_3 ~({\cal S}+1) ~(2{\cal S}+1)} ~,
\label{heff}
\eea
and ${\vec {\cal S}}_{\rm eff}$ denotes an object with spin ${\cal S}-1/2$. We
thus find a weak interaction between ${\vec {\cal S}}_{\rm eff}$ and all the 
sites labeled as $n=1$ in Fig. 3. [If the impurity has ${\cal S}=1/2$, the 
electron at $n=0$ forms a singlet with the impurity.] For ${\cal S} > 1/2$, 
we see that Eq. (\ref{heff}) has the same form as 
in Eqs. (\ref{hspin3}-\ref{jab1}), where the effective couplings 
$J_{\rm 1, eff} = 4 F_{\rm 1, eff} ~\sin^2 k_F$ and $J_{\rm 2, eff} = 4 
F_{\rm 2, eff} ~\sin^2 k_F$ are equal, negative and small. With these initial
conditions, we see from Eq. (\ref{disc}) and Fig. 1 that the Kondo couplings 
flow to the FP at $(J_{\rm 1, eff}, J_{\rm 2, eff}) = (0,0)$. 

We thus obtain a picture of the RG flows at both short and large length scales.
We start with the Griffiths $S$-matrix with certain values of the Kondo 
couplings, and finally end at the multi-channel FP of the disconnected 
$S$-matrix. But in this letter, we have restricted ourselves to weak 
inter-electron interactions, since we use perturbative methods to analyze the 
effects of $g_i$. Hence, the Luttinger parameter $K$ is restricted to be 
less than but close to unity. An interesting question to address is whether 
this analysis is true for strong inter-electron interactions when $K$ is 
much less than unity. In the two-wire case, it was shown that at strong 
interactions, it is the two-channel {\it antiferromagnetic} Kondo fixed point
which is stabilized for $K<1/2$\cite{fabrizio}. An equivalent analysis for 
$N$ wires is required\cite{uslong}. 

To summarize, we have studied the Kondo effect at a junction of $N$ quantum 
wires and find an interesting interplay of the Kondo logarithms and the 
TLL power laws. We find that the scaling of the Kondo couplings 
depends on the $S$-matrix at the junction. For the case of disconnected wires
and repulsive interactions, there is a range of Kondo couplings which flow
towards a multi-channel FP at $(J_1,J_2) = (0,0)$. At low temperatures,
we find spin-flip scattering processes with temperature dependences which are 
dictated by both the Kondo effect and the inter-electron interactions. 
It may be possible to observe such scatterings by placing a quantum dot with 
a spin at a junction of several wires with interacting electrons. 
At the fully connected or Griffiths $S$-matrix, we find that the Kondo 
couplings flow to a strong coupling FP, where their fate is decided
by a $1/J$ analysis. There is a range of initial conditions which again lead 
to the FP at $(J_{1, {\rm eff}}, J_{2, {\rm eff}})= (0,0)$.

\acknowledgments
SR thanks Y. Oreg and A. M. Finkel'stein for discussions. DS thanks the 
Department of Science and Technology, India for financial support under 
projects SR/FST/PSI-022/2000 and SP/S2/M-11/2000.

\end{document}